\documentclass[aps, prl, preprint, superscriptaddress]{revtex4-1}
\usepackage[english]{babel}
\usepackage{amsmath,amsthm}
\usepackage{amsfonts}
\usepackage{xspace}
\usepackage{bm}
\usepackage[squaren, thinqspace]{SIunits}
\usepackage{booktabs}
\usepackage{graphicx}

\theoremstyle{definition}

\theoremstyle{remark}
\begin{document}
\renewcommand{\degree}{\ensuremath{^\circ}\xspace}
\renewcommand{\Re}{\ensuremath{\mathrm{Re}} \xspace}
\renewcommand{\Im}{\ensuremath{\mathrm{Im}} \xspace}
\newcommand{\Hy}{\ensuremath{\bm{H} || \bm{y}}\xspace}
\newcommand{\Hx}{\ensuremath{\bm{H} || \bm{x}}\xspace}
\newcommand{\y}{\ensuremath{\bm{y}}\xspace}
\newcommand{\x}{\ensuremath{\bm{x}}\xspace}
\newcommand{\z}{\ensuremath{\bm{z}}\xspace}
\newcommand{\ii}{\ensuremath{\mathrm{i}}\xspace}
\newcommand{\Vdc}{\ensuremath{V_\mathrm{DC}}\xspace}
\newcommand{\Vsp}{\ensuremath{V_\mathrm{SP}}\xspace}
\newcommand{\Vvna}{\ensuremath{V_1}\xspace}
\newcommand{\gSpinMix}{\ensuremath{g_{\uparrow\!\downarrow}}\xspace}
\newcommand{\VISHEdc}{\ensuremath{V^\mathrm{dc}_\mathrm{iSHE}}\xspace}
\newcommand{\VISHEac}{\ensuremath{V_\mathrm{iSHE}}\xspace}
\newcommand{\Js}{\ensuremath{\bm{J}_\mathrm{s}}\xspace}
\newcommand{\mus}{\ensuremath{\bm{\mu}_\mathrm{s}}\xspace}
\newcommand{\Jc}{\ensuremath{\bm{J}_\mathrm{c}}\xspace}
\newcommand{\alphaSH}{\ensuremath{\Theta_{\mathrm{SH}}}\xspace}
\newcommand{\lambdaSD}{\ensuremath{\lambda_{\mathrm{SD}}}\xspace}
\newcommand{\tN}{\ensuremath{t_{\mathrm{N}}}\xspace}
\newcommand{\sigmaN}{\ensuremath{\sigma_{\mathrm{N}}}\xspace}
\newcommand{\tF}{\ensuremath{t_{\mathrm{F}}}\xspace}
\newcommand{\tPy}{\ensuremath{t_{\mathrm{Py}}}\xspace}
\newcommand{\sigmaF}{\ensuremath{\sigma_{\mathrm{F}}}\xspace}
\newcommand{\hrf}{\ensuremath{\bm{h}_\mathrm{mw}}\xspace}
\newcommand{\HexB}{\ensuremath{\bm{H}_0}\xspace}
\newcommand{\Hex}{\ensuremath{H_0}\xspace}
\newcommand{\Heff}{\ensuremath{\bm{H}_\mathrm{eff}}\xspace}
\newcommand{\M}{\ensuremath{\bm{M}}\xspace}
\newcommand{\my}{\ensuremath{m_y}\xspace}
\newcommand{\hy}{\ensuremath{h_y}\xspace}
\newcommand{\mz}{\ensuremath{m_z}\xspace}
\newcommand{\My}{\ensuremath{M_y}\xspace}
\newcommand{\Mz}{\ensuremath{M_z}\xspace}
\newcommand{\Ms}{\ensuremath{M_\mathrm{s}}\xspace}
\newcommand{\Vind}{\ensuremath{V_\mathrm{FMI}}\xspace}
\newcommand{\Eind}{\ensuremath{\bm{E}_\mathrm{ind}}\xspace}
\newcommand{\Eishe}{\ensuremath{\bm{E}_\mathrm{iSHE}}\xspace}
\newcommand{\muBohr}{\ensuremath{\mu_\mathrm{B}}\xspace}
\newcommand{\Hres}{\ensuremath{H_\mathrm{res}}\xspace}
\newcommand{\Ztot}{\ensuremath{Z}\xspace}
\newcommand{\Zishe}{\ensuremath{Z_\mathrm{iSHE}}\xspace}
\newcommand{\Zind}{\ensuremath{Z_\mathrm{FMI}}\xspace}
\newcommand{\Phifit}{\ensuremath{\phi_\mathrm{fit}}\xspace}
\newcommand{\Phiref}{\ensuremath{\phi_\mathrm{fit}^\mathrm{ref}}\xspace}
\newcommand{\Phiishe}{\ensuremath{\phi_\mathrm{iSHE}}\xspace}
\newcommand{\Phiind}{\ensuremath{\phi_\mathrm{FMI}}\xspace}
\newcommand{\Phitot}{\ensuremath{\phi}\xspace}
\newcommand{\Phiindy}{\ensuremath{\phi_\mathrm{FMI}^{y}}\xspace}
\newcommand{\chiT}{\ensuremath{\bm{\chi}}\xspace}
\newcommand{\PhiCoFe}{\ensuremath{\phi_\mathrm{CoFe}^{y}}\xspace}
\newcommand{\PhiPy}{\ensuremath{\phi_\mathrm{Py}^{y}}\xspace}

\title{Phase-sensitive detection of spin pumping via the ac inverse spin Hall effect}%

\author{Mathias Weiler}
\author{Justin M. Shaw}
\author{Hans T. Nembach}
\author{Thomas J. Silva}
\affiliation{Electromagnetics Division, National Institute of Standards and Technology, Boulder, CO, 80305}
\email{mathias.weiler@nist.gov}
\date{\today~~Contribution of NIST, not subject to copyright}%

\begin{abstract}
  An intriguing feature of spintronics~\cite{Zutic:2004} is the use of pure spin-currents to manipulate magnetization~\cite{Bader:2010}, e.g., spin-currents can switch magnetization in spin-torque MRAM~\cite{Chen:2010}, a next-generation DRAM alternative. Giant spin-currents via the spin Hall effect ~\cite{Dyakonov:1971,Hirsch:1999,Zhang:2000} greatly expand the technological opportunities~\cite{Liu:2012}.  Conversely, a ferromagnet/normal metal junction emits spin-currents under microwave excitation, i.e.~spin-pumping~\cite{Tserkovnyak:2002,Tserkovnyak:2005,Heinrich:2003}. While such spin-currents are modulated at the excitation frequency, there is also a non-linear, rectified component that is commonly detected using the corresponding inverse spin Hall effect (iSHE) dc voltage~\cite{Azevedo:2005,Saitoh:2006,Mosendz:2010a,Ando:2011}. However, the ac component should be more conducive for quantitative analysis, as it is up to two orders of magnitude larger and linear~\cite{Jiao:2013}. But any device that uses the ac iSHE is also sensitive to inductive signals via FaradayÕs Law and discrimination of the ac iSHE signal must rely on phase-sensitive measurements. We use the inductive signal as a reference for a quantitative measurement of the magnitude and phase of the ac iSHE.
\end{abstract}
\maketitle

%
\begin{figure}
  \centering
  \includegraphics{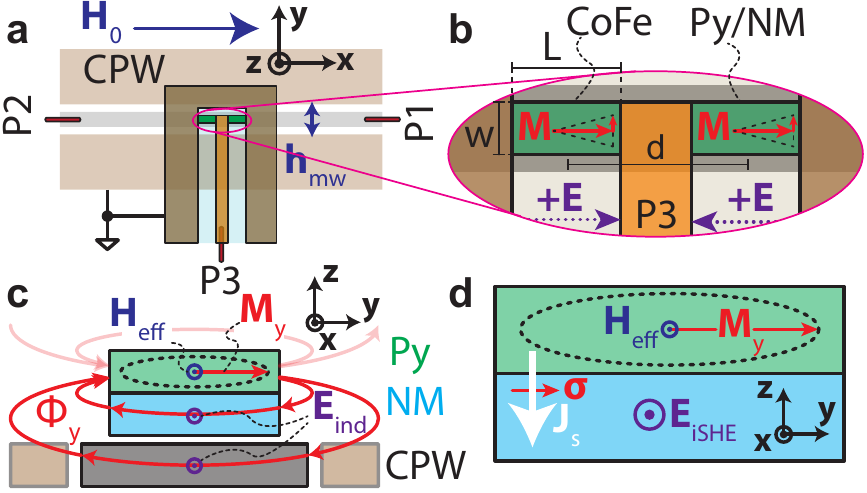}\\
  \caption{\textbf{Device and principles of inductive and ac iSHE signal generation.} \textbf{a} Schematic depiction of the excitation CPW (ports P1 and P2) and detection CPW (P3) placed at an angle of 90\degree. \textbf{b} Closeup of detection CPW.  \textbf{c} In the Py/NM tab, the time-varying flux $\Phi_y$ due to the dynamic magnetization component \My along the \y axis threads around the CPW center conductor and the NM and causes inductive signals \Eind along \x. \textbf{d} The dynamic spin current \Js due to \My gives rise to an ac electric field \Eishe along \x by virtue of the spin Hall effect.}\label{fig:device}
\end{figure}

Inductive voltages have long been exploited for the detection of magnetization dynamics~\cite{Silva:1999} and exhibit similar experimental signatures and magnitudes as the signals expected from the ac iSHE~\cite{Jiao:2013}. We use the three-terminal device depicted in Fig.~\ref{fig:device}a to separate ac iSHE and inductive signals by means of a phase-sensitive measurement technique. Application of an ac voltage to port 1 (P1) of the excitation coplanar waveguide (eCPW) generates a microwave magnetic field $\hrf\parallel\y$ above its center conductor. The detection coplanar waveguide (dCPW) at P3 is mounted at an angle of 90\degree and with an air gap of \unit{50}{\micro\meter} on top of the eCPW.  The dCPW is \unit{50}{\ohm} terminated by two rectangular thin-film tabs [$L\times w=\unit{(300\times100)}{\micro\meter\squared}$, center-to-center separation $d=\unit{325}{\micro\meter}$] as depicted in Fig.~\ref{fig:device}b. For all samples, the left tab is \unit{15}{\nano\meter} thick Co$_{90}$Fe$_{10}$ and the right tab is a Ni$_{81}$Fe$_{19}$ Permalloy (Py) thin film capped with various normal metal (NM) layers. Each tab has a dc resistance of approximately \unit{100}{\ohm}. Because the microwave termination for the dCPW is highly symmetric, the direct electromagnetic coupling from P1 to P3 is less than \unit{-28}{dB} for frequencies up to \unit{20}{\giga\hertz} (see SI). This allows us to employ a vector network analyzer (VNA) to directly measure both ferromagnetic induction (FMI) and ac iSHE signals without need for either external compensation circuits~\cite{Wei:2013} or nonlinear excitation schemes~\cite{Hahn:2013}.

For a static external magnetic field \HexB applied along the \x direction, parallel to the eCPW, the equilibrium magnetization \M points along the effective magnetic field $\Heff\approx\HexB$. The magnetization has dynamic components \Mz and \My as it precesses around \Heff with angular frequency $\omega$ in ferromagnetic resonance (FMR), as depicted schematically in Fig.~\ref{fig:device}b. Flux $\Phi_y$ due to the \My component threads around the center conductor of the eCPW and the NM of the Py/NM bilayer in the dCPW, as shown in Fig.~\ref{fig:device}c. According to Faraday's Law, $-d\Phi_y /d t$ gives rise to an electric field \Eind along \x in the Py/NM tab. The magnitude of the corresponding voltage \Vind is, to good approximation~\cite{Silva:1999}
\begin{equation}\label{eq:Vind}
\Vind=-i \omega \frac{\mu_0 L \tF}{2}\my \eta\;, 
\end{equation}
where $\mu_0$ is the vacuum permeability, $\My=\my e^{i \omega t}$, \tF is the ferromagnetic thin film thickness, and $0\leq \eta\leq1$ accounts for attenuation due to non-zero spacing between the Py and eCPW center conductor. It is important to note that magnetization dynamics thus cause inductive voltages in the eCPW and in the dCPW tabs in much the same way, with the only difference being the value of $\eta$.

\begin{figure}
  \centering
  \includegraphics{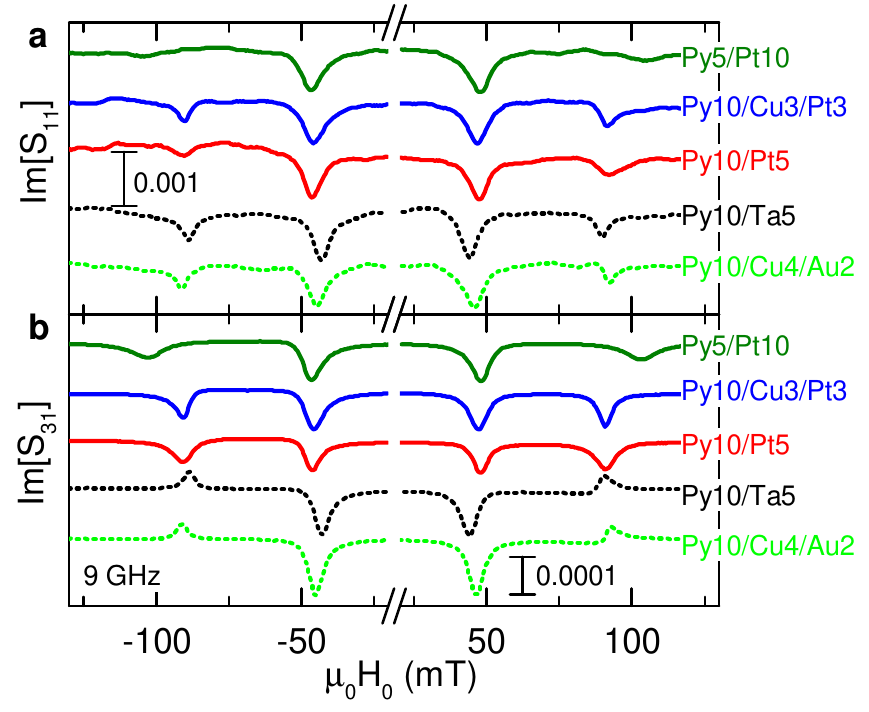}\\
  \caption{\textbf{Measured inductive and ac iSHE voltages.} \textbf{a} Measurements of $S_{11}$ vs.~\Hex (only imaginary part shown for clarity) which corresponds to the inductive signals induced in the eCPW. \textbf{b} Simultaneously acquired $S_{31}$ vs.~\Hex that has both inductive and ac iSHE contributions. The apparent change of sign of the Py/NM resonance for samples with Pt cap (solid lines, "dips") relative to those without Pt cap (dashed lines, "peaks") indicates a dominant non-inductive contribution in the samples with Pt cap. Integer numbers in the sample names denote nominal layer thicknesses in nm.}\label{fig:rawdata}
\end{figure}
To measure $\Vind$ in the eCPW, we measure the scattering parameter $S_{11}$ vs. \Hex  with the VNA at a fixed microwave frequency $f$. Figure~\ref{fig:rawdata}a shows data for $S_{11}$ vs.~\Hex obtained at $f=\unit{9}{\giga\hertz}$ for all samples. Note that $S_{11}$ is a complex quantity, but only the imaginary (absorptive) part is shown in Fig.~\ref{fig:rawdata}a. 

We first focus on the $S_{11}$ spectra obtained for the Py10/Cu4/Au2 sample. For either polarity of \Hex, two resonances are observed, one at $\mu_0|\Hres|\approx\unit{40}{\milli\tesla}$ and one at $\mu_0|\Hres|\approx\unit{80}{\milli\tesla}$. The dips at smaller absolute \Hres are due to the FMI-detection of the FMR of the CoFe tab, and the dips at larger field magnitudes are due to the FMR of the Py/NM tab, as verified by fitting of the data for \Hres to the Kittel equation (see SI). Both resonances are at the same phase relative to the excitation field \hrf and approximately symmetric with respect to inversion of the \Hex direction. This is in accordance with the detection of $\my=\chi_{yy}\hy$, where $\chi_{yy}$ is a diagonal component  of the magnetic susceptibility tensor $\chiT$ and $\chi_{yy}$ is even under external magnetic field inversion. The $S_{11}$ data remain qualitatively unchanged for all other samples, with the exception of Py5/Pt10, where the Py FMI signal is below the noise due to both the reduced ferromagnetic volume, and the spin-pumping-induced linewidth broadening, as discussed further below.

An $S_{31}$ measurement includes both the FMI signal of Eq.~\eqref{eq:Vind} and an iSHE signal due to ac spin pumping across the Py/NM interface. The basic idea for the ac iSHE signal generation is sketched in Fig.~\ref{fig:device}d: The precessing magnetization is damped in part by an ac spin current \Js pumped into the NM layer. \My gives rise to an ac electric field $\Eishe\propto\bm{\sigma}\times\Js$ due to the spin Hall effect, where $\bm{\sigma}\parallel\y$ is the spin-current polarization and $\Js\parallel\z$ is the direction of spin current flow. The magnitude of the ac iSHE voltage along \x is~\cite{Jiao:2013}
\begin{equation}\label{eq:Vishe}
\VISHEac=\frac{\gSpinMix}{2\pi}\;\omega e\alphaSH \lambdaSD\;\frac{\my}{\Ms} \;\frac{\tanh \left(\frac{\tN}{2\lambdaSD}\right)}{\tF \sigmaF+\tN \sigmaN} L\;, 
\end{equation}
where \gSpinMix is the effective interfacial mixing conductance with units of m$^{-2}$, $e$ is the electron charge, \alphaSH is the spin Hall angle of the normal metal, \lambdaSD is the spin diffusion length in the NM, \Ms is the saturation magnetization, \sigmaF is the conductivity of the FM, and \tN and \sigmaN are the thickness and conductivity of the NM, respectively. Because both \Vind and \VISHEac are proportional to $\my$, there is no qualitative difference between inductive and ac iSHE voltages other than a factor of $-i$, i.e., a $-90\degree$ phase shift. Furthermore, the ratio \VISHEac/\Vind is estimated to be in the order of unity for typical Pt/Py bilayers: using $\tF=\tN=\unit{10}{\nano\meter}$, $\sigmaF=\sigmaN=\unit{3\times10^6}{(\ohm\meter)^{-1}}$, $\Ms=\unit{800}{\kilo\ampere\per\meter}$, $\gSpinMix=\unit{3.5\times10^{19}}{\meter^{-2}}$, $\lambdaSD=\unit{1}{\nano\meter}$ and $\alphaSH=0.1$ as typical material parameters, we find $\VISHEac/\Vind\approx0.3$. Thus, reliable separation of ac iSHE and FMI signals requires \textit{phase-sensitive} detection.

In Fig.~\ref{fig:rawdata}b, we plot the imaginary part of $S_{31}$ acquired simultaneously with $S_{11}$. In $S_{31}$, signals of similar amplitude are observed at the resonance fields of both CoFe and Py for all samples. As there is no reasonable expectation for a large ac iSHE in CoFe or Py10/Cu4/Au2, we presume that both resonance signals are due to FMI solely~\footnote{The fact that an inductive signal is observed for a single CoFe layer is attributed to the non-uniform dynamic magnetization excitation through the film thickness.}~\cite{Maksymov:2013}. While the CoFe FMI signal is very similar for all samples and both orientations of \Hex, the shape of the Py/NM resonance changes from a peak for Py10/Ta5 and Py10/Cu4/Au2 (dashed lines) to a dip for the samples with Pt caps (solid lines)~\footnote{The apparent phase inversion of the Py10/Ta5 and Py10/Cu4/Au2 $S_{31}$ peaks relative to the $S_{11}$ signal is the result of the inverted polarity for the Py/NM tabs relative to the CoFe tab in the dCPW as indicated in Fig.~\ref{fig:device}b}. This is consistent with a phase shift of approximately 180\degree correlated with the presence of Pt in the NM stack. The behavior observed for Py10/Ta5 and Py10/Cu4/Au2 is consistent with the presumption that the ac iSHE signal is negligible in both cases.

\begin{figure}
  \centering
  \includegraphics{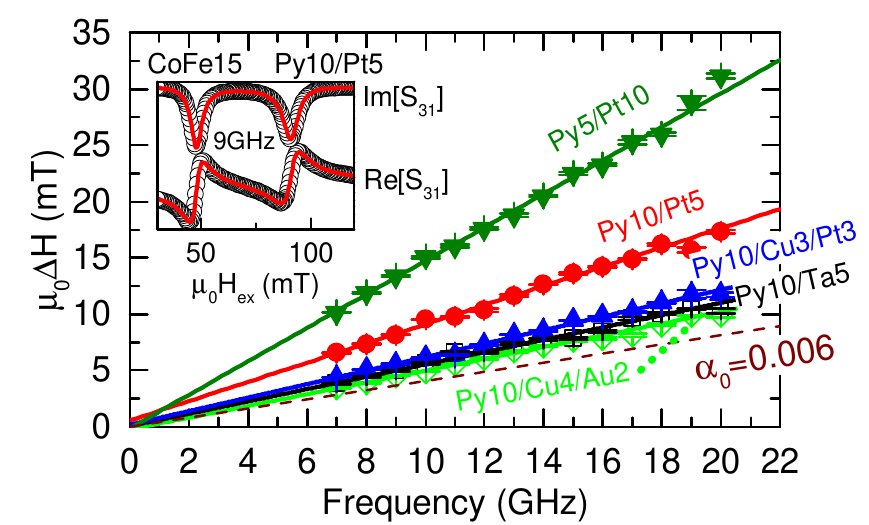}\\
  \caption{\textbf{Increased damping due to spin pumping}. Linear fits (solid lines) to the line width $\Delta H$ (symbols with error bars) of the Py/NM resonance extracted from $S_{31}$ measurements shows near zero inhomogeneous broadening (intercept of the linear fits) for all samples. The increase of damping (slope of the linear fits) relative to the intrinsic damping $\alpha_0=0.006$ (dashed line) is attributed entirely to spin pumping. }\label{fig:damping}
\end{figure}
To quantify the ac iSHE effect in our devices, we begin by fitting the $S_{31}$ data to a linear superposition of $\chi_{yy}$ for the two  magnetic susceptibilities $\bm{\chi}_\mathrm{CoFe}$ and $\bm{\chi}_\mathrm{Py}$. (See SI for details). An example of such a fit is shown in the inset of Fig.~\ref{fig:damping} for Py10/Pt5. We extract from the fits the resonance magnetic field $\Hres$ (see SI), the line width $\Delta H$, the magnitude $Z$, and the phase $\phi$ of the CoFe and Py resonances as a function of frequency for all samples. A linear fit of $\Delta H(f)$ as shown in Fig.~\ref{fig:damping} is  used to extract the total damping $\alpha$ and, thereby \gSpinMix . Results are tabulated in the SI. 

While the extraction of $\Hres$ and $\Delta H$ from susceptibility measurements is a standard procedure~\cite{Kalarickal:2006}, quantification of ac iSHE signals rests on the analysis of  $Z$ and $\phi$, as all other parameters are common to both ac iSHE and FMI.  For purely inductive and pure ac iSHE signals we expect
\begin{equation}\label{eq:Z}
\Zind e^{i\Phiind}=\varepsilon\frac{\Vind(\Hres)}{\Vvna\chi_{yy}(\Hres)}\;\;,\;\; \Zishe e^{i \Phiishe}=\varepsilon\frac{\VISHEac(\Hres)}{\Vvna\chi_{yy}(\Hres)}\:,
\end{equation}
respectively. \Vvna is the ac voltage applied at P1, and the dimensionless factor $0\leq\varepsilon\leq1$ accounts for losses in the dCPW. Both \Zind and \Zishe are normalized to the magnetic susceptibility such that they are otherwise independent of the FMR response. If $S_{31}$ can be characterized as a linear superposition of FMI and ac iSHE responses, we can use
\begin{equation}\label{eq:Ztot}
Z e^{i\phi}=\Zishe e^{i\Phiishe}+\Zind e^{i\Phiind}
\end{equation}
to deduce the magnitude and phase of the ac iSHE and FMI signals. As detailed in the SI, \Phitot is referenced to the resonance phase of the CoFe tab.
\begin{figure*}
  \centering
  \includegraphics{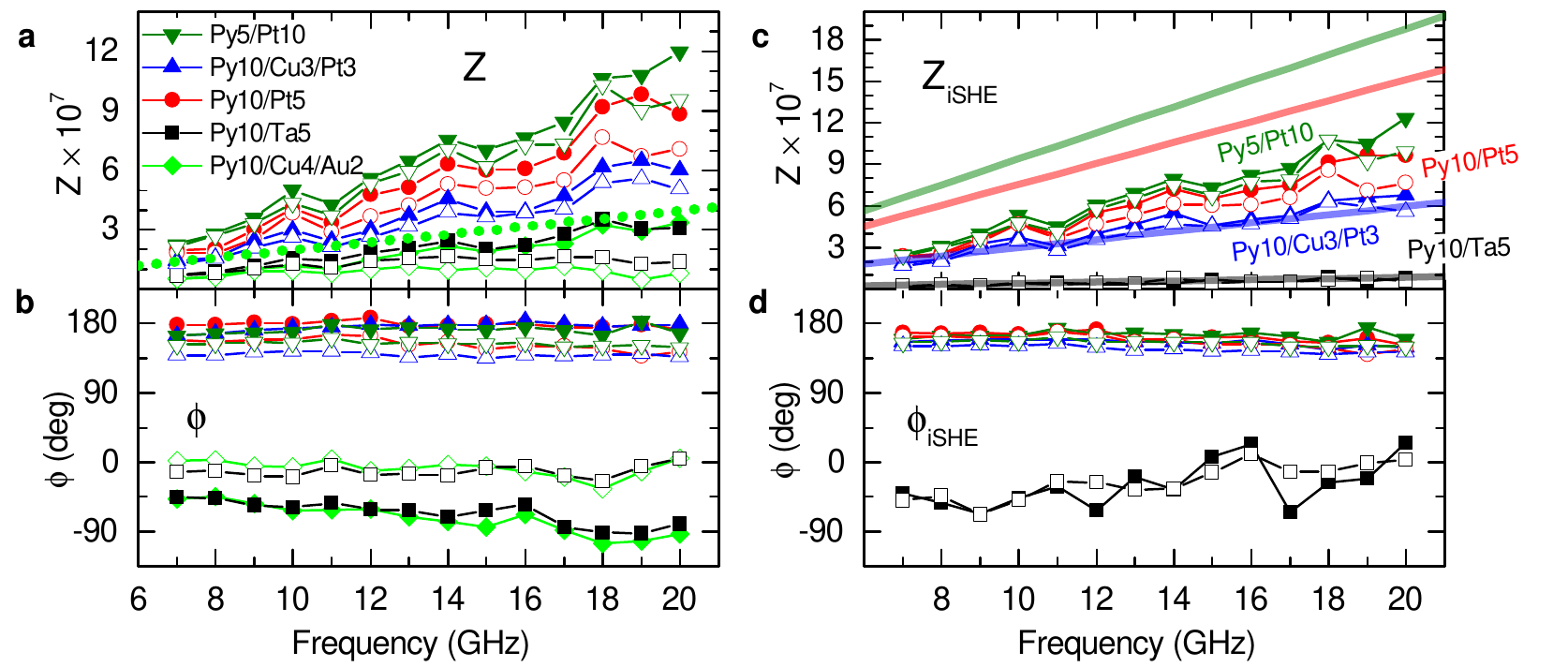}\\
  \caption{\textbf{Magnitude and phase of inductive and ac iSHE signals.} \textbf{a} The fitted magnitude \Ztot for the Py/NM resonances of all samples. Solid symbols correspond to $\HexB \parallel +\x$ and open symbols to $\HexB \parallel -\x$. The green dotted line is an estimate for a purely inductive signal for a \unit{10}{\nano\meter}-thick Py film according to Eq.~\eqref{eq:Z}. \textbf{b} The fitted phase relative to the phase of the CoFe resonance for each sample. \textbf{c} The extracted magnitude of the ac iSHE signal obtained by subtraction of the inductive signal contribution. Solid lines are calculations based on Eq.~\eqref{eq:Z}. \textbf{d} The extracted phase of the ac iSHE signal relative to that of the CoFe resonance is approximately 160\degree at all frequencies for samples with Pt caps. The sample with a Ta cap shows an approximately 180\degree inverted phase attributed to the sign-change of the spin Hall angle. }\label{fig:acishe}
\end{figure*}
We plot the extracted \Ztot and \Phitot as a function of frequency for all investigated Py/NM bilayers and both \HexB polarities in Fig.~\ref{fig:acishe}a and~b, respectively. \Ztot is largest for the samples with a Pt cap and maximum for the Py5/Pt10 sample. This is not expected if the signals were purely FMI. From Eq.~\eqref{eq:Vind}, one would expect $Z\propto\tF$ such that the Py5/Pt10 signal would be half of that of Py10/Pt5. 

Regarding the phase, $\Phitot=0\degree$ is expected for a pure FMI. Instead, we find $\Phitot\approx 160\degree$ for all samples capped with Pt, while $-90\degree\lesssim\Phitot\lesssim0\degree$ for the  samples without Pt. The large phase shift caused by inclusion of a Pt cap is indicative of an additional non-inductive signal source due to the presence of Pt.
 
 Under the presumption that the additional signal is the result of the ac iSHE, we extract the ac iSHE contribution from the variation of \Ztot and \Phitot between the various samples. The signal from the Py10/Cu4/Au2 sample is effectively due solely to FMI, and we assume that the same FMI signal is present in all Py/NM stacks, except for the Py5/Pt10 sample, where we scale the magnitude of the inductive signal by a factor of one half. By use of Eq.~\eqref{eq:Ztot}, we obtain \Zishe and \Phiishe, shown in Figs.~\ref{fig:acishe}c and~d, respectively. Both \Zishe and \Phiishe are even under \HexB inversion, consistent with the symmetry of $\chi_{yy}$. In contrast, \Zind and \Phiind exhibit an asymmetry under inversion of \Hex, which is not expected for a signal entirely due to \My (see green diamonds in Fig.~\ref{fig:acishe}a and ~\ref{fig:acishe}b). As we subtract the measured \Zind(Py10/Cu4/Au2) from \Ztot to obtain \Zishe, a full quantitative understanding of the FMI signal is however not required.
 
 The $180\degree$ phase difference between samples with Pt and Ta cap in Fig.~\ref{fig:acishe}d is in accordance with the sign change of \alphaSH from Pt to Ta~\cite{Liu:2012}. Equations~\eqref{eq:Vind} and~\eqref{eq:Vishe} predict a 270\degree (90\degree) phase difference of inductive to ac iSHE signals with positive (negative) \alphaSH. As we find  $\Phiishe\approx160\degree$ for the samples with a Pt cap and $\Phiishe\approx-20\degree$ for the Py/Ta sample, we observe a $\approx110\degree$ lag in the ac iSHE phase. This suggests that \alphaSH in metals is in actuality a complex quantity at microwave frequencies. Dispersion for spin accumulation via the SHE has been previously observed in semiconductors~\cite{Stern:2008}, but a retardation effect for charge-/spin-current interconversion is surprising since it is generally assumed that electron momentum scattering sets the relevant time scale.

 We now estimate the expected FMI and ac iSHE signals by use of Eqs.~\eqref{eq:Vind} to~\eqref{eq:Z}. We obtain $\chi_{yy}$ directly from fitting the spectra, the damping and \gSpinMix from the linear fits to the data in Fig.~\ref{fig:damping} to Landau-Lifshitz theory, and a fit of \Hres to the Kittel equation (see SI) yields $g\approx2.1$ for the Py resonances of all samples. We use \sigmaN and \sigmaF from 4-probe dc resistance measurements of bare films.  The only uncertain parameters are $\varepsilon$, \lambdaSD, and \alphaSH. We first assume that \Vind from Eq.~\eqref{eq:Vind} quantitatively accounts for the inductive signal. Then, \Zind for Py/Cu4/Au2 and \Zishe for Py10/Cu3/Pt3 and Py10/Ta5 can be modeled by use of $\varepsilon=0.08$, $\lambdaSD^\mathrm{Pt}=\unit{1}{\nano\meter}$~\cite{Boone:2013} and $\alphaSH^\mathrm{Pt}=0.9$, $\lambdaSD^\mathrm{Ta}=\unit{1}{\nano\meter}$~\cite{Boone:2013} and $\alphaSH^\mathrm{Ta}=-0.08$ in Fig.~\ref{fig:acishe}a (dotted line) and~\ref{fig:acishe}c (solid lines), respectively. This estimates the upper limit for $\alphaSH$ in Pt and Ta for our samples. However, additional attenuation of the inductive signal for Py/Cu4/Au2 is entirely possible as a result of a non-uniform dynamic magnetization depth profile due to eddy currents~\cite{Maksymov:2013} and to shunting of the FMI signal by the NM layer that affects the source compliance. A lower limit for $\alphaSH$ is obtained by assuming zero losses ($\varepsilon=1$) resulting in $\alphaSH^\mathrm{Pt}=0.072$ and $\alphaSH^\mathrm{Ta}=-0.006$. Thus, \Zishe is within the range expected from reported $\alphaSH$ of Pt~\cite{Liu:2011,Weiler:2013} and Ta~\cite{Morota:2011, Liu:2012, Wang:2013}. Eq.~\eqref{eq:Vishe} overestimates \Zishe obtained for the two Py/Pt samples by a factor of 2, which may be caused by interfacial spin flip~\cite{Nguyen:2013, Rojas-Sanchez:2013}.

\section{Supplementary Information}

\subsection{Device assembly and characterization}
\begin{figure}
  \centering
  \includegraphics{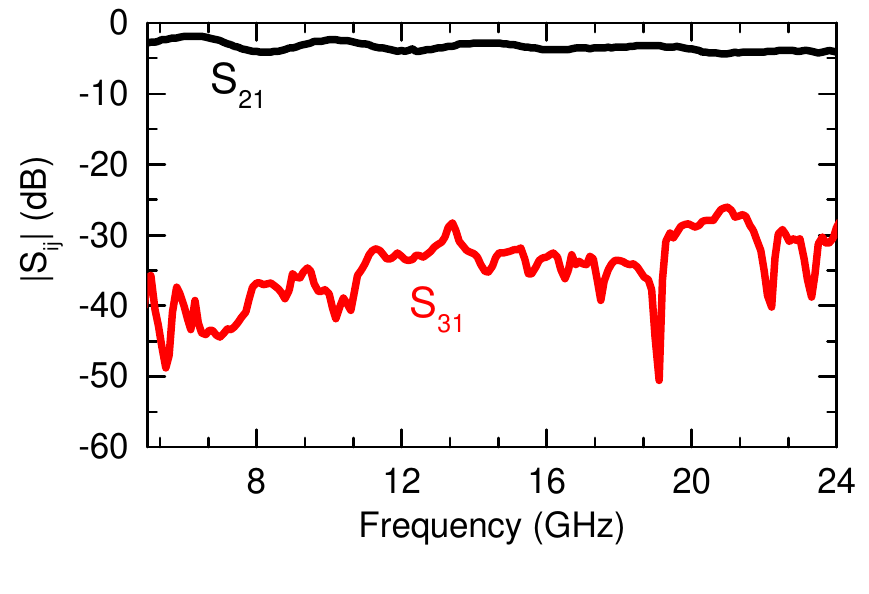}\\
  \caption{\textbf{S-parameter Measurements for the Py10/Cu4/Au2 sample.} The transmission magnitude $|S_{21}|$ (black line) through the excitation CPW is above \unit{-5}{dB} for $f\leq\unit{24}{\giga\hertz}$, indicating high bandwidth and low loss of the eCPW. The spurious microwave crosstalk $|S_{31}|$ (red line) from the eCPW to the dCPW is less than \unit{-28}{dB} in this frequency range. }\label{fig:Sparams}
\end{figure}

The three terminal devices used for the measurements were assembled from two CPWs (the detection and excitation CPW) that were mounted at a fixed angle of $(90\pm5)\degree$ with an air gap of $\delta\approx\unit{50}{\micro\meter}$ by an aluminum sample holder designed specifically for this purpose. The excitation CPW (eCPW) connecting P1 and P2 is made from a \unit{35}{\micro\meter} thick Cu film on an \unit{200}{\micro\meter} thick alumina substrate mounted on an aluminum base plate. The eCPW center conductor width is $w_\mathrm{CPW}=\unit{150}{\micro\meter}$ and the gap to the ground planes is \unit{108}{\micro\meter}. The detection CPWs (dCPW)  are fabricated via thin-film sputter-deposition of Cu(\unit{180}{\nano\meter})/Au(\unit{20}{\nano\meter}) bilayers  on \unit{400}{\micro\meter} thick, double-side polished, optically transparent sapphire substrates. The dCPW has a center conductor width of \unit{25}{\micro\meter} and a gap of \unit{25}{\micro\meter}. The dCPW was designed to be \unit{50}{\ohm} impedance matched by use of a 3D planar electromagnetic field solver (Sonnetsoftware Sonnet 14~\footnote{Certain commercial instruments are identified to specify the experimental study adequately. This does not imply endorsement by NIST or that the instruments are the best available for the purpose}). The ferromagnetic thin film patches were deposited on the sapphire substrate before deposition of the dCPW. Connections to the three ports of the device are made using \unit{2.4}{\milli\meter} end launch connectors. Alignment of the CoFe and Ni$_{80}$Fe$_{20}$/NM (Py/NM) tabs with the center conductor of the eCPW was performed with the aid of an optical microscope.

We characterized the microwave properties of the assembled devices using a calibrated vector network analyzer (VNA). A 2-port calibration was performed to P1 and P3 of the device using an electronic calibration kit. In Fig.~\ref{fig:Sparams}, we provide an example of S-parameter data obtained from the mounted Py10/Cu4/Au2 device without application of an external magnetic field. Low insertion loss of the eCPW is observed [$|S_{21}|$ (black line)]. The $<\unit{5}{dB}$ losses in $|S_{21}|$ in this frequency range are attributed mostly to the connection of the end launches to the eCPW and signal attenuation due to the ground planes of the dCPW at distance $\delta$ to the eCPW. The crosstalk between port 1 and 3 of the device [$|S_{31}|$ (red line)] is below \unit{-28}{dB} in the frequency range of interest. This is expected for the symmetric arrangement of CoFe and Py/NM layers on our dCPW that results in destructive interference of any non-ferrous inductive cross-coupling between the eCPW and dCPW. The strong suppression of field-independent coupling between the waveguides is reflected in the high signal-to-noise ratio observed in $S_{31}$-measurement in Fig.~2 in the main text.

\subsection{Fitting of the data}
\begin{figure}
  \centering
  \includegraphics{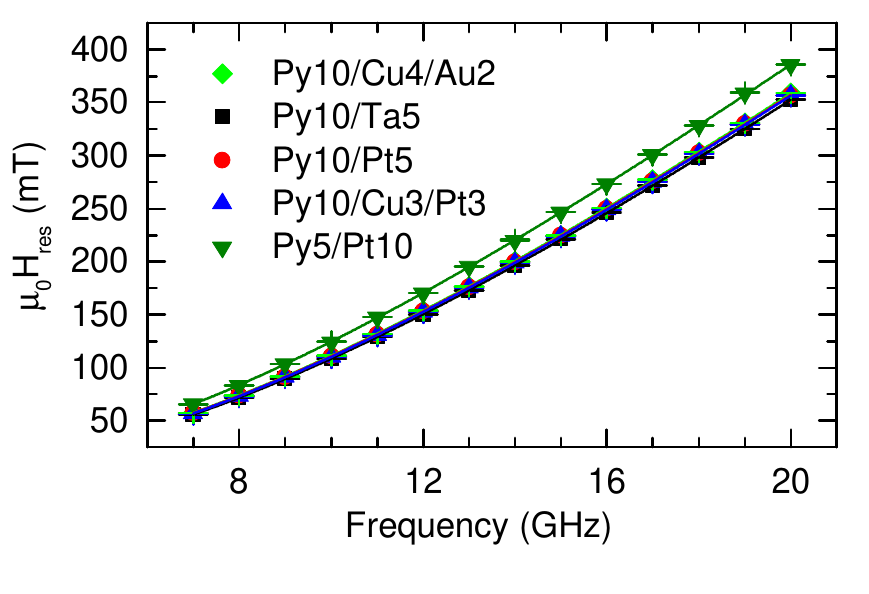}\\
  \caption{\textbf{Measured \Hres and Kittel fits.} Measured resonance field \Hres for the Py/NM resonances of all samples (symbols with fit errors) and corresponding Kittel fits according to Eq.~\eqref{eq:Kittel} (lines). }\label{fig:Hres}
\end{figure}
\begin{table*}
\begin{tabular}{lccccc}
\colrule  
  Sample & $M_\mathrm{eff}$ (kA/m) & $ g$ & $\mu_0 \Delta H_0$ (mT) & $\alpha$ & \gSpinMix ($10^{19}$ m$^{-2}$) \\ 
\colrule    
Py10/Cu4/Au2 & $730\pm5$ & $2.112\pm0.005$ & $-0.2\pm0.1$ & $0.0077\pm0.0001$ & $0.88$\\
Py10/Ta5 & $742\pm7$ & $2.121\pm0.007$ & $0.1\pm0.2$ & $0.0081\pm0.0002$ & $1.08$\\
Py10/Pt5 & $733\pm5$ & $2.115\pm0.005$ & $0.6\pm0.1$ & $0.0126\pm0.0002$ & $3.39$\\
Py10/Cu3/Pt3 & $739\pm2$ & $2.110\pm0.002$ & $0.2\pm0.1$ & $0.0088\pm0.0002$ & $1.46$\\
Py5/Pt10 & $624\pm9$ & $2.128\pm0.01$ & $-0.1\pm0.3$ & $0.0221\pm0.0004$ & $4.11$\\
\botrule   
\end{tabular}
\caption{Fitted parameters from the Py/NM resonance for the samples used in this study. Numbers indicate layer thickness in nm. The effective spin mixing conductance \gSpinMix was estimated using Eq.~\eqref{eq:gSpinMix} assuming $\alpha_0=0.006$.}
\label{tab:samples}
\end{table*}
All $S_{31}$ spectra were fitted to the superposition of the complex $\chi_{yy}$ components of the two magnetic susceptibilities $\chi^\mathrm{Py/NM}$ and $\chi^\mathrm{CoFe}$ with~\cite{Dreher:2012}
\begin{equation}\label{eq:chi}
\chi^\mathrm{(j)}=\frac{\mu_0 M_\mathrm{eff,(j)}}{D_{(j)}}\left(\begin{array}{cc}
\mu_0 \Hex+\frac{i \omega \alpha_{(j)}}{\gamma_{(j)}} & +\frac{i \omega}{\gamma_{(j)}}\\
-\frac{i \omega}{\gamma_{(j)}} & \mu_0 (M_\mathrm{eff,(j)}+\Hex) +\frac{i \omega \alpha_{(j)}}{\gamma_{(j)}}
\end{array}\right)\;,
\end{equation}
where $(j)=$Py/NM or CoFe and with
\begin{equation}\label{eq:D}
D=\left(\mu_0 (M_\mathrm{eff, (j)}+\Hex) +\frac{i \omega \alpha_{(j)}}{\gamma_{(j)}}\right)\left(\mu_0 \Hex+\frac{i \omega \alpha_{(j)}}{\gamma_{(j)}}\right)-\left(\frac{\omega}{\gamma_{(j)}}\right)^2\;,
\end{equation}
\begin{equation}\label{eq:alpha}
\alpha_{(j)}=\mu_0 \Delta H_{(j)} \frac{\gamma_{(j)}}{2 \omega}\;,
\end{equation}
and $M_\mathrm{eff}$ from
\begin{equation}\label{eq:Kittel}
\omega=\mu_0\gamma\sqrt{\Hres(\Hres+M_\mathrm{eff})}\;.
\end{equation}
Here, $\gamma=g\muBohr/\hbar$ is the gyromagnetic ratio with the spectroscopic g-factor $g$,  and Eq.~\eqref{eq:Kittel} is the Kittel equation for in-plane geometry with $\omega=2\pi f$. Equation~\eqref{eq:alpha} does not take inhomogeneous broadening into account. This does not influence our results, as we recover the inhomogeneous broadening from the fitted $\Delta H$ as discussed below. The fits to the complex $S_{31}$ data are performed by a Levenberg-Marquardt optimization of
\begin{equation}\label{eq:Fits}
S_{31}(\Hex)= \sum_{j} Z_{(j)} e^{i\phi_{(j)}} \chi_{yy}^{(j)}(\Hex)+C_1+C_2\Hex\;.
\end{equation}
Here, $C_1$ and $C_2$ are complex offset and slope, respectively. They account for background and drift in $S_{31}(\Hex)$. In addition to $C_1$ and $C_2$, we obtain one set of fit parameters $Z$, $\phi$, $\Delta H$ and $H_\mathrm{res}$ for each resonance. 

As detailed in Ref.~\cite{Nembach:2011}, for the fit of $S_{31}$ to Eq.~\eqref{eq:Fits}, we set $g=2$. The $g$-factor is recovered together with $M_\mathrm{eff}$ in a second step, where we fit the \Hres vs. $f$ data to $\Hres(f)$ extracted from Eq.~\eqref{eq:Kittel}. Data and corresponding fits for \Hres are shown in Fig.~\ref{fig:Hres}.  We then obtain the damping $\alpha$ and the inhomogeneous broadening $\Delta H_0$ from a linear fit to the $\Delta H$ data shown in Fig.~3 in the main text to
\begin{equation}\label{eq:gilbert}
\mu_0\Delta H=\mu_0 \Delta H_0+\frac{2\omega \alpha}{\gamma}\;,
\end{equation} 
where we use the value for $g$ extracted from the Kittel fit in the previous step.

We now estimate the effective spin mixing conductance \gSpinMix by~\cite{Tserkovnyak:2005}
\begin{equation}\label{eq:gSpinMix}
\gSpinMix=\left(\alpha-\alpha_0\right)\frac{4\pi \Ms \tF}{\hbar \gamma}\;, 
\end{equation}
where we use $\alpha_0=0.006$ determined from a conventional VNA-FMR measurement of an unpatterned Py thin film and $M_\mathrm{s}=\unit{800}{\kilo\ampere\per\meter}$. The obtained material parameters for all samples are summarized in Table~\ref{tab:samples}. We observe very similar values of $g\approx2.1$ and $\Delta H_0\approx0$ for all samples, indicative of high uniformity of the sample magnetization, anisotropy and external magnetic field. Within fitting error, $M_\mathrm{eff}$ is the same for all Py10/NM samples. The reduction of $M_\mathrm{eff}$ for the Py5/Pt10 layer is attributed to interfacial anisotropy~\cite{Shaw:2013}. The extracted effective mixing conductances \gSpinMix are within the range of expectations for these combinations of materials~\cite{Boone:2013, Jiao:2013}. \gSpinMix is largest for the samples where the Py layer is in direct contact with a Pt layer, while insertion of a Cu layer significantly reduces \gSpinMix, consistent with experimental findings~\cite{Weiler:2013} and theoretical estimates\cite{Jiao:2013}. Small effective mixing conductances for the samples with Ta and Au cap are mainly attributed to the effect of spin backflow~\cite{Jiao:2013}, which is large for materials with either a long spin diffusion length (Au) or low conductivity (Ta)~\cite{Boone:2013}.

\subsection{Reference resonator}
\begin{figure*}
  \centering
  \includegraphics{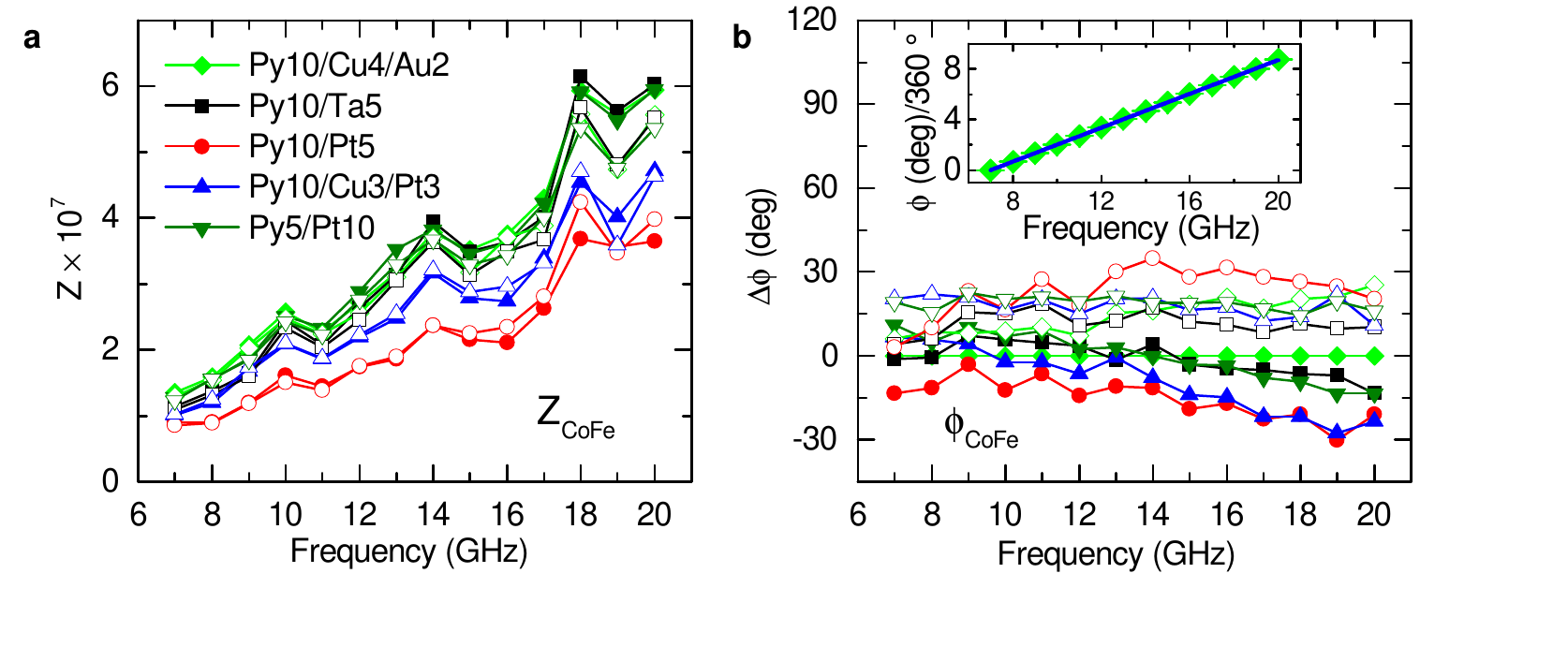}\\
  \caption{\textbf{Measured $Z$ and $\phi$ for the CoFe tabs.} \textbf{a} $Z_\mathrm{CoFe}$ extracted from fits to $S_{31}$ using Eq.~\eqref{eq:Fits}. Solid symbols are for $\Hex>0$ and open symbols for $\Hex<0$. Fit errors are smaller than symbol size. \textbf{b} Corresponding $\phi_\mathrm{CoFe}$ relative to $\phi_\mathrm{CoFe}^\mathrm{ref}$ obtained for the Py10/Cu4/Au2 sample with $\Hex>0$.  The inset shows the raw $\phi_\mathrm{CoFe}^\mathrm{ref}$ data (green diamonds) together with a linear fit (blue line).}\label{fig:CoFe}
\end{figure*}
Fitting results for $Z_\mathrm{CoFe}$ and $\phi_\mathrm{CoFe}$ for all samples are shown in Fig.~\ref{fig:CoFe}a and~\ref{fig:CoFe}b, respectively. $Z_\mathrm{CoFe}$ is independent of the polarity of \Hex within good approximation. Variations of $Z_\mathrm{CoFe}$ among the different samples are mainly attributed to changing impedance with the different Py/NM stacks. We use the phase $\phi_\mathrm{CoFe}^\mathrm{ref}$ obtained for the Py10/Cu4/Au2 sample at positive \Hex as a reference and plot $\Delta\phi=\phi_\mathrm{CoFe}-\phi_\mathrm{CoFe}^\mathrm{ref}$. The phases $\Delta \phi$ shown in Fig.~\ref{fig:CoFe}b are determined from fits to the CoFe resonances of all samples. All fitted phases $\phi_\mathrm{CoFe}$ increase linearly with frequency, as demonstrated for the CoFe resonance of the Py10/Cu4/Au2 sample in the inset of Fig.~\ref{fig:CoFe}b. This is expected due to the non-zero electrical length of the signal paths from P1 to P3 which results in a frequency dependent phase $\phi=360\degree f l/c$ with $c=c_0/\sqrt{\varepsilon_r}$, the relative permittivity $\varepsilon_r\approx10$ of the waveguides and the speed of light in vacuum $c_0$. The linear fit shown in the inset of Fig.~\ref{fig:CoFe}b has a slope of $241\degree/\mathrm{GHz}$ corresponding to an electrical length of $l\approx\unit{6.4}{\centi\meter}$. This is in agreement with the physical length of the signal path from P1 to P3. 
 
From Fig.~\ref{fig:CoFe}b we observe that $\Delta \phi$ varies by up to 60\degree under \Hex inversion. This is attributed to contributions of $d\Mz/dt$ component to the FMI. The corresponding component $\chi_{zy}$ of the magnetic susceptibility tensor is odd under \HexB inversion and 90\degree out-of-phase with $\chi_{yy}$. Under this presumption we obtain the phase of the FMI of the \My component of the CoFe tab as $\PhiCoFe=\left[\phi_\mathrm{CoFe}(\Hex>0)+\phi_\mathrm{CoFe}(\Hex<0)\right]/2$. The phase \PhiPy of the FMI due to the \My component of the Py/NM tabs is then calculated (including the effect of different electrical lengths for signals passing from P1 through either CoFe or Py/NM tabs to P3) as  $\PhiPy=\PhiCoFe-360\degree f d/c+180\degree$. Here, $d=\unit{325}{\micro\meter}$ is the center-to-center separation of the Py/NM and CoFe tab. The additional factor of 180\degree is obtained because of the symmetry of the arrangement of CoFe and Py/NM tabs on the dCPW. (See main text). In the main text, the presented values for $\phi$ are referenced to the calculated value $\PhiPy$, i.e. $\phi=\phi_\mathrm{Py}-\PhiPy$, where $\phi_\mathrm{Py}$ is obtained from fits of the $S_{31}$ spectra to Eq.~\eqref{eq:Fits}.

\subsection{Calculation of driving field}
In order to estimate \Vind and \VISHEac from the equations given in the main text, $\my=\chi_{yy}\hy$ needs to be known. While we obtain $\chi_{yy}$ from our fits as discussed above, we calculate $\hy$ by use of the Karlqvist equations~\cite{Mallinson:1993} as
\begin{equation}\label{eq:hrf}
\hy= \frac{V_1}{Z_0 w_\mathrm{CPW}\pi} \arctan\left(\frac{w_\mathrm{CPW}}{2 \delta}\right)  \zeta\;, 
\end{equation}
with $\zeta=0.5$ accounting for the non uniform current distribution in the center conductor that reduces \hy at the position of the CoFe and Py/NM tabs with width $w<w_\mathrm{CPW}$ as detailed in Ref.~\cite{Neudecker:2006}.
%


\end{document}